\documentclass[conference]{IEEEtran}
\ifCLASSOPTIONcompsoc
  \usepackage[nocompress]{cite}
\else
  \usepackage{cite}
\fi

\IEEEoverridecommandlockouts
\usepackage{cite}
\usepackage{amsmath,amssymb,amsfonts}
\usepackage{algorithmic}
\usepackage{graphicx}
\usepackage{textcomp}
\usepackage{xcolor}
\usepackage{lipsum}
\usepackage{mathtools}
\usepackage{cuted}
\usepackage{subcaption}
\usepackage{soul}
\usepackage{multirow}
\usepackage{hyperref}

\def\BibTeX{{\rm B\kern-.05em{\sc i\kern-.025em b}\kern-.08em
    T\kern-.1667em\lower.7ex\hbox{E}\kern-.125emX}}
    
\begin{document}

\title{Stepwise Migration of a Monolith to a Microservices Architecture: \\ Performance and Migration Effort Evaluation}

\author{Diogo~Faustino,
        Nuno~Gonçalves,
        Manuel~Portela,
        António~Rito~Silva,
\IEEEcompsocitemizethanks{\IEEEcompsocthanksitem D. Faustino, N. Gonçalves and A. Rito Silva are with DPSS - INESC-ID, Instituto Superior Técnico, University of Lisbon, Lisbon, Portugal.
E-mail: \{diogofaustino, nuno.m.bagulho.goncalves, rito.silva\}@tecnico.ulisboa.pt
\IEEEcompsocthanksitem M. Portela is with CLP - Centre for Portuguese Literature, University of Coimbra, Coimbra, Portugal.
E-mail: mportela@fl.uc.pt
}
}

\IEEEtitleabstractindextext{%
\begin{abstract}
The agility inherent to today's business promotes the definition of software architectures where the business entities are decoupled into modules and/or services. However, there are advantages in having a rich domain model, where domain entities are tightly connected, because it fosters an initial quick development. On the other hand, the split of the business logic into modules and/or services, its encapsulation through well-defined interfaces and the introduction of inter-service communication introduces a cost in terms of performance. In this paper we analyze the stepwise migrating of a monolith, using a rich domain object, into a microservice architecture, where a modular monolith architecture is used as an intermediate step. The impact on the migration effort and on performance is measured for both steps. Current state of the art analyses the migration of monolith systems to a microservices architecture, but we observed that migration effort and performance issues are already significant in the migration to a modular monolith. Therefore, a clear distinction is established for each one of the steps, which may inform software architects on the planning of the migration of monolith systems. In particular, the trade-offs of doing all the migration process or just migrating to a modular monolith. 
\end{abstract}

\begin{IEEEkeywords}
Domain-driven design, Modular architecture, Migration effort, Performance evaluation, Microservices.
\end{IEEEkeywords}}

\maketitle

\IEEEdisplaynontitleabstractindextext

%
\IEEEpeerreviewmaketitle

\section{Introduction}
\label{sec:intro}

Domain-driven design~\cite{Evans03} advocates the division of a large domain model into several independent bounded contexts to split a large development team into several small, and business focused, teams and foster an agile software development. These bounded contexts can be implemented as modules, each one of them with a well-defined interface, to further decouple the teams by reducing the number of required interactions between them. This modularization is also suggested as an intermediate step of the migration of a monolith to a microservices architecture~\cite{Richardson19}.

The correct identification of what should be the responsibilities associated with each module is not trivial, and has to be done through several refactoring steps~\cite{Opdyke93,Evans03,Fowler18}. These refactoring operations can be more easily performed in a strongly connected domain model where the business logic is scattered among the domain entities, a rich domain model, and in the absence of interfaces between the domain entities~\cite{Haywood17}. So, projects often start with a single domain model, because in the first development phases the domain model is not completely understood. Premature modularization adds a cost to development, because of the need to refactor modules' interfaces, since the initial design does not capture the correct abstractions.

The use of interfaces between modules requires the transformation of data between them to encapsulate their domain models, anticorruption layers~\cite{Evans03}, which implies significant changes do the domain model, and have impact on the performance of the system. Therefore, the process of modularizing a monolith system, and further migrating it to a microservices architecture~\cite{Fowler14}, has to address these problems.

The migration into a microservice architecture encapsulates the bounded contexts into independent processes that become cohesive services, capable of being individually deployed and isolated through an API that encapsulates and provides the anti-corruption layers. Therefore, a first decomposition of a monolith into modules can provide the groundwork for implementing microservices, separating the aspects of modularization from those of distribution. Only, the latter has to address inter-process communication and decentralized data management.

Research has been done on the comparison of the performance quality between a monolith system and its correspondent implementation using a microservices architecture, but these results are sometimes contradictory, e.g.~\cite{Villamizar15,Ueda16}, address different characteristics of a microservices system, e.g.~\cite{Joy15,Al-Debagy18}, or are evaluated using simple systems, e.g.~\cite{Tapia20}. Furthermore, in these studies, it was not necessary to redesign the monolith functionalities in order to be migrated to the microservices architecture.

In this paper, we describe the stepwise migration process of a large monolith system into a microservices architecture, removing circular dependencies between modules, and adapting each of these modules into independent microservices with well defined REST APIs, while focusing on the migration effort and the performance overheads. We also describe the types of refactoring that were applied to the domain model and measure their impact on the migration effort. Regarding performance, we compare the performance of the three implementations: monolith, modular monolith and microservices. Additionally, we describe the architectural tactics that were applied to improve the performance of both the modular monolith and the microservice architecture.

The results of this paper also provide a contribution to the methods and techniques for migrating a monolith to a microservices architecture, since the modular monolith can be used has an intermediate step in the process of migration to a microservices architecture. For instance, our results show a significant impact on latency, even in the absence of the remote invocations between modules. Additionally, it has the advantage that by having an intermediate step, our analysis can provide insights on the monolith architectural migration to a microservices architecture that are independent of the technological aspects associated with the implementation of a microservices architecture.

In the next section we describe the stepwise migration process of a monolith to a micorservices architecture, using a modular monolith as an intermediate step. Then, in section~\ref{sec:evaluation} the migration process is evaluated in terms of migration cost and performance, and the results are discussed in section~\ref{sec:discussion}, which highlights the migration trade-offs and the lessons learned. Section~\ref{sec:related_work} presents related work and section~\ref{sec:conclusion} the conclusions.

\section{Migration}
\label{sec:modular_monolith}

The migration of a monolith with a rich and complex domain into loosely decoupled subdomains is a stepwise process that first migrates to a modular monolith and then to a set of microservices. 

\subsection{Modular Monolith}

The advantage of the modular monolith is that it conciliates small teams, one for each part of the domain model, with transactional execution of the monolith functionalities, because all parts of the domain model are stored in the same database~\cite{Haywood17}. Therefore, it is necessary to describe how the domain is partitioned into several disconnected parts and how the modules interact.

\subsubsection{Module Interfaces}

To migrate the monolith to a set of modules the domain model is split into several non-interrelated domain models. The interactions between these smaller domain models are carried out through the module interfaces. The modular monolith defines two types of interactions between modules, a uses interaction, where a module uses another by invoking its interface, and a notification interaction, where a module notifies of its changes the modules that subscribe to the notifications. The uses interaction defines a dependence in which the module that uses requires the used module to be present~\cite{documenting_software_architecture}. In this type of interaction, the module that is used has a \textit{provides interface} that match a \textit{requires interface} defined in the client module.
The notification interaction allows a module to inform other modules about the occurrence of a change without creating a dependence between them~\cite{documenting_software_architecture}. This is achieved by sending events that contain the information. The module that notifies has a \textit{publish interface} while the module that is notified has a \textit{subscribe interface}, and they only need to agree on the event structure. In the uses interaction, between a requires and a provides interface, data is transferred in the form of Data Transfer Objects (\textit{DTO})~\cite{Fowler02}, which contain information on the module domain model in an inter-module shared format that preserves the encapsulation of the used module domain entities. This contributes to modules' loose-coupling. 
The notification interactions define the events a module publishes and that are subscribed by other modules. These events contain information about what occurred in the module and which may be relevant to subscribing modules. Usually, the events contain minimal information and, if necessary, the subscribing modules may afterwards start a uses interaction to obtain more information about the state of the module that sent the notification. Note that, the uses interaction establishes a flow of control from the use modules to the used modules, and the notification interface defines the inverse flow of control. However, the dependencies only occur in the uses flow of control direction, because the correct functioning of the module that notifies does not require the correct implementation of the module that is notified~\cite{documenting_software_architecture}. 

\subsubsection{Domain Model Refactoring}

The modularization of a monolith requires the decompositon of its domain models into different modules, such that each one of the modules does not inter-depend on a shared data repository~\cite{Haywood17}.

To partition the monolith domain model, it is necessary to consider the relationships between the domain entities that will belong to different modules. These relationships need to be replaced by invocations to the modules interface. Additionally, since circular dependencies are not allowed in a modular monolith, it is also necessary to decide what will be the relations between the modules, and implement their interactions using uses and notification interfaces. The implementation of these interactions requires the definition of \textit{DTOs} and events. 

Another aspect that needs to be addressed, when modularizing a monolith, is the presence of god classes~\cite{Olbrich10}, classes that tend to centralize the intelligence of the system, in the monolith. Therefore, it is necessary to split these classes into several classes, one for each partition of the domain model. This refactoring requires the identification of which methods of the god class access which part of the domain model, and moves them to the corresponding new singleton. This task is trivial, except when the methods access more than a single module, but, this case, is like the above problem of refactoring inter-module relationships. Therefore, the complexity of refactoring of the god classes reduces to the complexity associated with refactoring that removes inter-module relationships.

In a rich domain model two types of relations between entities are possible: associations, which allow for the specification of various types of multiplicities, such as one-to-one and one-to-many, and inheritance. When a superclass has subclasses that will belong to different modules, it is necessary to add all methods that are implemented in the parent class to the child classes. Similarly to the refactoring of god classes, the complexity of refactoring inheritance hierarchies reduces to the complexity of refactoring the inter-modules associations they use.

The types of refactoring of a rich domain model reduce to the case of removing an association between modules. To do this refactoring, it is necessary to keep a unique identifier of the domain entity of the used module in the module that uses it. This unique identifier is used by the use module to obtain a \textit{DTO} with information of the entity, and for the used module to notify a change of its state by publishing an event. Note that, the returned \textit{DTO} may contain unique identifiers of other domain entities, due to the indirect relationships between domain entities.

\subsubsection{Performance Optimizations}

The definition of several modules in the modular monolith raises performance problems associated with the inter-module communication and the need of data transformations to isolate each module's domain model. During domain model refactorization it is necessary to evaluate the performance of the migrated functionalities. If they show poor performance, the number of invocations between modules have to the reduced. This is achieved by redesigning the inter-module interactions into coarse-grained invocations. Note that, monoliths are often implemented using fine-grained object-oriented interactions. 

The following performance optimizations are applied:
\begin{itemize}
    \item Associate a database access index to the unique identifiers of domain entities exported to other modules, due to the frequent invocations to obtain a \textit{DTO} object of the domain entity, given its unique identifier;
    \item Whenever a \textit{DTO} is generated, by default, it is loaded with the values for all its attributes, which work as a cache, to reduce further inter-module invocations to obtain each one of its values. This introduces a memory overhead, because some of the attribute values may not be necessary in the context of that particular interaction. However, overall, after systematically applying this tactic, the performance of the system improved;
    \item In a few cases, after performance evaluation, it is necessary to have larger \textit{DTO} objects, which, additionally to the attribute values, also contain the information associated with several interrelated domain entities. For instance, a \textit{DTO} of a list of \textit{DTO}s may reduce the number of inter-module invocations when a module is interacting with a collection of domain entities in another module.
\end{itemize}

\subsection{Microservices}

The modular monolith defines the units of modularity of the microservices system. Therefore, the modules defined in the modular monolith are converted into services with little effort, which allows the developers to focus on more intricate aspects of a microservice architecture like inter-service communication, decentralized data management and data consistency.

\subsubsection{Microservices API}

The microservice architecture defines different types of inter-service communication, synchronous communication, where a service requests the information through an API and awaits its response, and asynchronous communication, which can undergo between multiple variants, depending on the requirements of the architecture, but mainly allows for services to share information without any dependencies. The synchronous communication presents a similar behaviour to the uses interaction where the services maintain the dependencies and request information through the network without redesigning the features of the service. This communication performs a typical request/response interaction using a technology built around HTTP like REST and replaces the uses interaction in a transparent manner to the behaviour of the application. The asynchronous communication allows to adapt the notification interaction while maintaining the behaviour of the service and avoiding circular dependencies between the services. This is achieved through an event-based communication where a publisher service is responsible for publishing events to all subscribed services through a message broker, responsible for managing the events and publishing them. Additionally, this communication plays an important role on the consistency of the information, resulting in eventual consistency between the databases of the services in the absence of distributed transactions.

\subsubsection{Interfaces Refactoring}

The focus of the migration of the modular monolith into a microservice is on the interfaces of the services. Both, the uses and notification interactions, have to be adapted into the appropriate type of communication. This requires a redesign of the provides/requires interface to implement synchronous communication, and the publisher/subscriber service to introduce the message broker and implement the asynchronous communication. 

The refactoring of the uses interactions is straightforward. The provides interface of the services are implemented as public endpoints which are invoked by the requires interface of a service through HTTP requests. The refactoring process consisted of mapping the publicly available methods from the interface to an unique URI and to an appropriate request type, GET, POST, PUT, and DELETE, according to the actions of the method in the information. On the other hand, the requires interface needs to match the changes to the provides interface and implement a remote access through the defined URL, with the appropriate mapping in terms of parameters and type of request.

An additional requirement related to the provides interface in the microservice architecture is the compatibility of the \textit{DTOs} to the serialization required in a remote invocation. Generally, this is a trivial process since the \textit{DTOs} have a simple structure and mostly carry domain information, however when carrying non-serializable information, this can require an additional effort in order to adapt the \textit{DTOs} without affecting the behaviour of the application. 

The refactoring of notification interactions introduces a message broker to achieve event-driven communication between the microservices. In general, the introduction of a message broker is not a complex task and mostly requires a publish and listen method for, respectively, the publisher and subscriber interfaces that communicates with the broker. Another aspect that needs to be addressed concerns the consistency of the information in a decentralized data architecture with asynchronous communication to notify services of relevant changes. Therefore, the microservice architecture has to consider eventual consistency between the databases, without which it can result in information inconsistency.

\subsubsection{Performance Optimization}

Similarly to the modular monolith, the definition of several microservices and the introduction of synchronous communication in the architecture raises performance problems associated with the number of remote invocations per functionality and the amount of information transferred in \textit{DTOs}. The following optimizations are applied:

\begin{itemize}
    \item Implementation of caches in the microservices for faster information retrieval speed and to reduce the number of remote invocations for faster performance. Some of these caches are created upon service creation and the data is preserved until the services stop their execution, this is very effective when the chosen data remains immutable throughout the execution;
    \item When necessary, \textit{DTOs} containing the information of several interrelated domain entities have to be defined to reduce the number of remote invocations, though it increases the amount of information sent in a single request. 
\end{itemize}

\section{Evaluation}
\label{sec:evaluation}

A large monolith was migrated to a microservices architecture to evaluate the development effort and the impact on performance of the stepwise migration process. 

The \textit{LdoD Archive} monolith\footnote{https://ldod.uc.pt/} is a digital archive for digital humanities having a rich domain model. It offers features like searching, browsing and viewing the original text fragments, different variations of the text fragments, as well as different editions of a book. It also provides a way for users to create their own (virtual) editions of the book, in which they can add, remove and order the fragments as they wish. Other features include a recommendation feature, that defines a proximity measure between different fragments according to a set of criteria, and a game feature, which implements a serious game where users tag text fragments.

The monolithic application is implemented in Java using the Spring-Boot framework\footnote{https://spring.io/} and an Object-Relational Mapper (ORM) to manage the domain model's persistence. The domain model has 71 domain entities and 81 bidirectional relations between domain entities, which resulted in a strongly coupled rich domain model. This solution provides high reusability, because the business logic is split among the domain entities and can easily be reused, and also due to the bidirectional relations that facilitate the navigation in the domain model. In total, the monolith has 25.862 lines of Java code, 20.039 lines of JSP code and 7.721 lines of JavaScript code.

The modular monolith implements the same features as the monolith. These features are implemented as a set of modules that apply the uses and notification interactions to preserve the dependencies between features: \texttt{Text}, which represents the fragments and their expert editions; \texttt{User}, which provides users registration, authentication, and access authorization; \texttt{Virtual}, which allows the definitions of virtual editions by end users and requires the \texttt{Text} and \texttt{User} modules; \texttt{Recommendation}, which calculates similarity distances between the fragments, given a set of criteria, such as date and tf-idf (Term Frequency Inverse Document Frequency)~\cite{Salton88}, between text fragments and uses the \texttt{Virtual} and \texttt{Text} modules; and \texttt{Game}, which implements a tagging serious game to classify the fragments in a virtual edition, and uses the \texttt{Virtual} module.

\subsection{Migration Effort}

The migration effort depends on the number of relationships between the new modules' domain entities and on the number of interfaces between modules. 

In what concerns the domain entities, although a domain entity may not have a direct relationship with a domain entity of another module, it may receive it as a parameter of an invocation. This is the dependency problem addressed by the Demeter Law~\cite{Lienberherr89}. So, these indirect relationships also need to be dealt with during the refactoring process.

\begin{figure*}
    \centering
    \includegraphics[width=1.0\textwidth]{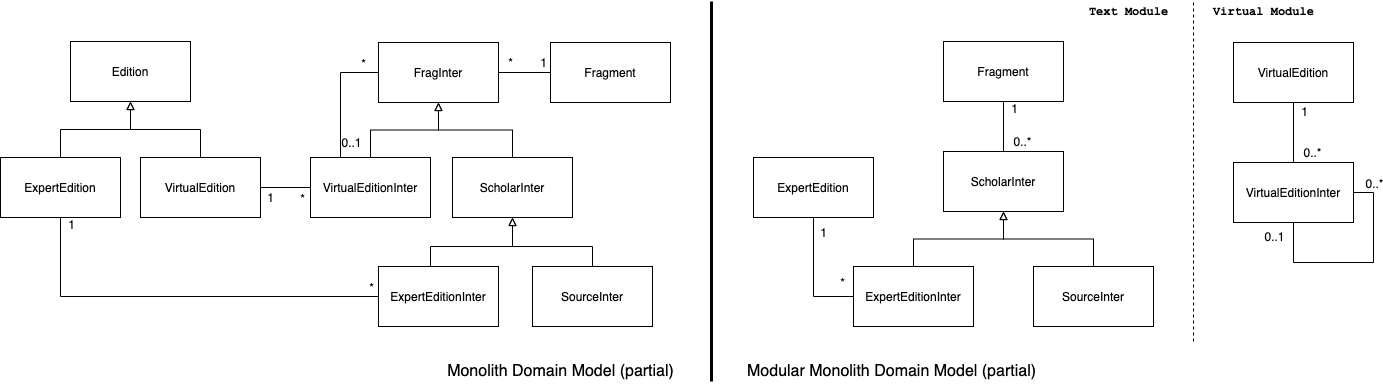}
    \caption{Domain model refactoring (partial)}
    \label{fig:relations}
\end{figure*}

Figure \ref{fig:relations} shows a refactoring that illustrates most of the found situations. The refactoring was applied to part of the monolith domain model, where abstract superclasses \texttt{Edition} and \texttt{FragInter} are removed and their subclasses split between \texttt{Text} and \texttt{Virtual} modules. The association between \texttt{FragInter} and \texttt{Fragment} is preserved in the \texttt{Text} module in entity \texttt{ScholarInter}, but in the \texttt{Virtual} module the fragment will be represented by its unique identification to be stored in\textit{VirtualEditionInter} . When necessary, the \texttt{Virtual} module can request the information of a fragment, given its unique identifier, and receive the corresponding \texttt{FragmentDto} object. Additionally, when a \texttt{Fragment} is deleted, the \texttt{Text} module emits an event that is subscribed by the \texttt{Virtual} module in order to delete the corresponding \texttt{VirtualEditionInter} entities. A similar refactoring is applied to the association between \texttt{FragInter} and \texttt{VirtualEditionInter} which results in a reflexive association in entity \texttt{VirtualEditionInter} together of a unique identifier of a \texttt{ScholarInter} entity.

\begin{table*}[]
\resizebox{\textwidth}{!}{%
\begin{tabular}{c|c|c|c|c|c|}
\cline{2-6}
                                                                  & \textbf{Text} & \textbf{User} & \textbf{Virtual} & \textbf{Recomendation} & \textbf{Game} \\ \hline
\multicolumn{1}{|c|}{\textbf{Modified Entities/Total Entities}}   & 23/42 (55\%)  & 2/5 (40\%)    & 14/17 (82\%)     & 1/2 (50\%)             & 3/5 (60\%)    \\ \hline
\multicolumn{1}{|c|}{\textbf{Defined DTOs/Total Entities}}        & 10/42 (24\%)  & 1/5 (20\%)    & 7/17 (41\%)      & 0/2 (0\%)              & 0/5 (0\%)     \\ \hline
\multicolumn{1}{|c|}{\textbf{Modified Relations/Total Relations}} & 6/38 (15\%)   & 4/7 (57\%)    & 7/26 (27\%)      & 0/2 (0\%)              & 0/6 (0\%)     \\ \hline
\multicolumn{1}{|c|}{\textbf{Removed Relations/Total Relations}}  & 2/38 (5\%)    & 0/7 (0\%)     & 3/26 (12\%)      & 2/2 (100\%)            & 5/6 (83\%)    \\ \hline
\end{tabular}
}
\caption{Impact of the refactoring in the domain model}
\label{tab:refactoring-stats-domain}
\end{table*}

Table~\ref{tab:refactoring-stats-domain} presents the impact of the domain model refactorings in terms of domain entities, and their relationships, of the different modules. It can be observed that there is a big impact on the domain models, where more than 50\% of the domain entities had to be changed, and in the case of the \texttt{Virtual} module this value reaches 82\%. This corresponds to a considerable effort associated with the migration from the monolith to the modular monolith. A similar situation can be observed in the percentage of \textit{DTOs} that had to be defined, given the total number of domain entities, which has impact on the changes to the modules that use these \textit{DTOs}, and in particular in the domain entities that interacted, before the refactorization, with the original domain entities. 

Note that, while the number of domain entities modified in the \texttt{Text} module might appear higher than it should be, since it does not depend on any other module, the majority of the changes were relatively small and focused mainly on the module's publish interface. It is necessary to remove code that interacts with entities that are located in other modules, and emit events instead. For instance, the \texttt{remove} method of the \texttt{FragInter} entity, that is moved to the \texttt{ScholarInter} subclass. Originally, this entity had a direct relationship to a set of \texttt{VirtualEditionInter} instances, allowing it to directly trigger the removal of those instances when it is removed. Since this relation was removed due to connecting two separate modules, this code was replaced with a call to the event interface to notify all modules of the removal of an instance of \texttt{ScholarInter} (\texttt{FragInter}).

In what concerns the impact on associations, two situations can occur: (1) modified relations, which correspond to associations where one of the involved domain entities changes, but not the overall meaning and purpose of the association; (2) removed associations, which correspond to associations between domain entities of different modules that have become represented by the domain entities unique identifiers. Considering the refactorings in Figure~\ref{fig:relations}, where the association between \texttt{FragInter} and \texttt{Fragment} is initially refactored into an association between \texttt{ScholarInter} and \texttt{Fragment} and another between \texttt{VirtualEditionInter} and \texttt{Fragment}, the former new association is modified, and the latter is removed.

It can be observed in Table~\ref{tab:refactoring-stats-domain} that there is a smaller impact of the refactoring on the associations, because only the associations between domain entities belonging to different modules are impacted, while more domain entities can be impacted due to the indirect relationships. However, the percentage of changed associations is higher if a module has less domain entities, because most of them will be in the interface with the other modules. Therefore, in the \texttt{Text} module a high number of relations are fully contained within the module, and it leads to a low percentage of modified relations when compared to other modules that required modifications. On the other hand, the \texttt{Recommendation} module is the outlier among the modules. The fact that all its relations are removed is the result of having only two domain entities, which had associations with domain entities belonging to other modules.

\begin{table*}[]
\resizebox{\textwidth}{!}{%
\begin{tabular}{c|c|c|cc|cc|cc|}
\cline{2-9}
                                                      & \textbf{Text} & \textbf{User} & \multicolumn{2}{c|}{\textbf{Virtual}}              & \multicolumn{2}{c|}{\textbf{Recommendation}}             & \multicolumn{2}{c|}{\textbf{Game}}                \\ \cline{2-9} 
                                                      & \textbf{Pro}  & \textbf{Pro}  & \multicolumn{1}{c|}{\textbf{Pro}}   & \textbf{Req} & \multicolumn{1}{c|}{\textbf{Pro}} & \textbf{Req} & \multicolumn{1}{c|}{\textbf{Pro}}  & \textbf{Req} \\ \hline
\multicolumn{1}{|c|}{\textbf{Modified/Total Methods}} & 91/109 (83\%) & 43/47 (91\%)  & \multicolumn{1}{c|}{152/155 (98\%)} & 6/6 (100\%)  & \multicolumn{1}{c|}{6/6 (100\%)}  & 7/7 (100\%)  & \multicolumn{1}{c|}{14/14 (100\%)} & 9/9 (100\%)  \\ \hline
\multicolumn{1}{|c|}{\textbf{New Methods}}            & 16            & 2             & \multicolumn{1}{c|}{2}              & 18           & \multicolumn{1}{c|}{1}            & 0            & \multicolumn{1}{c|}{0}             & 2            \\ \hline
\multicolumn{1}{|c|}{\textbf{Modified/Total DTOs}}    & 6/15 (40\%)   & 2/3 (67\%)    & \multicolumn{2}{c|}{9/14 (64\%)}                   & \multicolumn{2}{c|}{3/11 (27\%)}                 & \multicolumn{2}{c|}{2/6 (33\%)}                   \\ \hline
\end{tabular}
}
\caption{Impact of the refactoring in the provides interfaces (P) and requires interfaces (R) and \textit{DTOs} from each of the microservices in terms of refactored methods and \textit{DTOs}. \textit{Text} and \textit{User} modules do not have required interfaces because they do not use other modules.}
\label{tab:refactoring-stats-micro}
\end{table*}

The refactorings to a microservice architecture are focused on two different major aspects of the modular architecture, the provides and requires interface and the corresponding \textit{DTOs} from each service. Table~\ref{tab:refactoring-stats-micro} presents the impact of refactoring the uses interaction into synchronous communication using a REST API. It can be observed a significantly high percentages of modified methods where at least 83\% of the methods had to be mapped. This may look like a significant migration effort, but these changes were simple changes to implement, consisting of repetitive mapping procedures.  

However, a few incompatibilities were detected that directly affected the refactoring cost which should be taken into consideration while developing the interfaces in the modular monolith for a smoother migration. For instance, an incompatibility was found in methods of the provides interface that required multiple complex objects as different parameters, which was not supported by the HTTP protocol due to the body of a HTTP request only being parsed once. In these cases, the definition of additional serializable \textit{DTOs} that wrap the parameters were required in these methods. Therefore, this is a simple solution with a low effort to implement. That is why there are more \textit{DTOs} than in the modular monolith.

Another incompatibility was the presence of non-serializable information being transferred between the modules that could not be transferred in a remote invocation of a service. This behaviour proved to have a more serious consequence, that not only affected the API but also the features from a service. An example of this incompatibility was found in the \textit{Recommendation} microservice where the methods utilized a non-serializable data structure to transfer the information in the \textit{DTOs}. A way to address this situation is by implementing a custom serializable configuration for that type of information, which would allow the serialization into a JSON format that could be sent between the services and be applicable to every usage throughout the interfaces. This is an efficient solution that allows the reuse of the information in the \textit{DTOs} and maintains the refactoring in the scope of the API, but the effort depends on the object. On the other hand, if the serialization cannot be achieved, it is necessary to individually address their usages and adapt how the information is transferred between the services while maintaining the behaviour of the functionality that depends on it. Therefore, the quality of the interfaces in the modular monolith have an effect on the migration effort and it is important to keep the information transferred between the modules simple for an efficient migration.

\subsection{Performance}

To evaluate the impact that the new architecture has on the performance of the system, we performed performance tests on the previously described features of the application, using JMeter\footnote{https://jmeter.apache.org/}. The testing was done with the application running inside Docker containers on a dedicated machine with an Intel I7 6 cores, 16GB of RAM and a 1TB of SSD.

Four functionalities were selected for the analysis. They were chosen due to their impact in terms of: the number of the domain entities they interact with, the number of modules/services accessed by the functionality, and the amount of processing required by the functionality.

The \textit{Source Listing} functionality presents the listing of the archive sources, where a source corresponds to a physical source document. When using this functionality, the end user obtains all the information about each one of the 754 sources, such as date, dimensions, and type of ink used in the document. The implementation of this functionality is done through an interaction between the front-end and \texttt{Text} modules/services. The modular monolith implementation was optimized to have a single invocation between the modules and a composed \textit{DTO} object is returned, which contains the list of all \texttt{FragmentDTO}s where each \texttt{FragmentDTO} embeds its list of \texttt{SourceDTO}s. 

The \textit{Fragment Listing} functionality is implemented in the modular monolith and microservice as an interaction between the front-end and \textit{Text} to present the list of all fragments. There are 720 text fragments in the archive. For each fragment the information of several interpretations are presented, and each fragment can have 2 to 7 interpretations. The front-end does a total of 5 invocations to the \textit{Text} module. Only one of the invocations required an additional optimization, in which the front-end obtains the list of \texttt{FragmentDTO}s, and each fragment contains its \texttt{ScholarInterDTO}s, both \texttt{ExpertEditionDTO}s and \texttt{SourceInterDTO}s. 

The \textit{Interpretation View} functionality presents an interpretation of a text fragment. Its implementation in the modular monolith and microservice requires several interactions between the front-end and the \textit{Text}, \textit{User} and \textit{Virtual} modules/services. To do the performance test it was chosen an interpretation of a virtual edition that includes several different domain entity types. It was not possible to optimize the implementation of the functionality using a coarse-grained invocation. On the other hand, the amount of information retrieved in each inter-module interaction is small.

The \textit{Assisted Ordering} functionality orders the fragments according to a set of criteria, such as date and text similarity (tf-idf). This ordering requires more than 250 000 fragment comparisons (considering for instance a virtual edition with 720 fragments, we get $720*719/2=258\ 840$ comparisons between fragments), and each comparison requires information about the fragment for up to 4 criteria. For its implementation the front-end accesses 4 back-end modules/services: \textit{Text}, \textit{User}, \textit{Virtual} and \textit{Recommendation}. Because this functionality repeatedly interacts through the same set of data, the information about the fragments is cached in the monolith implementation, to improve performance. The modular monolith implementation also uses this cache while the microservice architecture implemented two additional caches due to performance deterioration caused by remote invocations of the same set of data.

\begin{table*}[]
\resizebox{\textwidth}{!}{%
\begin{tabular}{c|c|c|c|c|c|}
\cline{2-6}
                                                   & \textbf{Text} & \textbf{User} & \textbf{Virtual} & \textbf{Recommendation} & \textbf{Game} \\ \hline
\multicolumn{1}{|c|}{\textbf{Source Listing}}      & 8/42 (19\%)   & 0/5 (0\%)     & 0/17 (0\%)       & 0/2 (0\%)               & 0/5 (0\%)     \\ \hline
\multicolumn{1}{|c|}{\textbf{Fragment Listing}}    & 9/42 (21\%)   & 0/5 (0\%)     & 0/17 (0\%)       & 0/2 (0\%)               & 0/5 (0\%)     \\ \hline
\multicolumn{1}{|c|}{\textbf{Interpretation View}} & 22/42 (52\%)  & 1/5 (20\%)    & 6/17 (35\%)      & 0/2 (0\%)               & 0/5 (0\%)     \\ \hline
\multicolumn{1}{|c|}{\textbf{Assisted Ordering}}   & 21/42 (50\%)  & 1/5 (20\%)    & 3/17 (18\%)      & 1/2 (50\%)              & 0/5 (0\%)     \\ \hline
\end{tabular}
}
\caption{Coverage of the domain entities by each of the functionalities}
\label{tab:refactoring-stats-func}
\end{table*}

Table~\ref{tab:refactoring-stats-func} presents the number of domain entity types, for each module, that a functionality accesses.

For each functionality, a test case was designed to compare the performance for all three architectures. Each test case run simulates a user that sequentially submits 50 requests, after a first request to warm up the caches. The test cases are run for two different loads of the database, for 100 and 720 fragments, respectively. The results are shown in Table \ref{tab:res10-2-monolith-modular}. 

\begin{table*}[]
\resizebox{\textwidth}{!}{%
\begin{tabular}{ll|c|c|c|c|c|c|}
\cline{3-8}
                                                                                                              & \multicolumn{1}{c|}{\textbf{}} & \textbf{Monolith} & \textbf{Modular} & \textbf{Microservices} & \textbf{\begin{tabular}[c]{@{}c@{}}Variation\\ Monolith \\ Modular\end{tabular}} & \textbf{\begin{tabular}[c]{@{}c@{}}Variation\\ Modular \\ Microservices\end{tabular}} & \textbf{\begin{tabular}[c]{@{}c@{}}Variation\\ Monolith \\ Microservices\end{tabular}} \\ \hline
\multicolumn{1}{|l|}{\multirow{2}{*}{\textbf{\begin{tabular}[c]{@{}l@{}}Source\\ Listing\end{tabular}}}}      & \textbf{Latency}               & 17/61             & 22/629           & 438/4317               & 29\%/931\%                                                                       & 1891\%/586\%                                                                          & 2476\%/6977\%                                                                          \\ \cline{2-8} 
\multicolumn{1}{|l|}{}                                                                                        & \textbf{Throughput}            & 55.1/16.2         & 43.7/1.6         & 2.3/0.23               & -21\%/-90\%                                                                      & -95\%/-86\%                                                                           & -96\%/-99\%                                                                            \\ \hline
\multicolumn{1}{|l|}{\multirow{2}{*}{\textbf{\begin{tabular}[c]{@{}l@{}}Fragment\\ Listing\end{tabular}}}}    & \textbf{Latency}               & 103/350           & 61/1095          & 725/5715               & -41\%/213\%                                                                      & 1089\%/422\%                                                                          & 604\%/1533\%                                                                           \\ \cline{2-8} 
\multicolumn{1}{|l|}{}                                                                                        & \textbf{Throughput}            & 9.6/2.9           & 16.1/0.91        & 1.4/0.18               & 68\%/-69\%                                                                       & -91\%/-80\%                                                                           & -85\%/-94\%                                                                            \\ \hline
\multicolumn{1}{|l|}{\multirow{2}{*}{\textbf{\begin{tabular}[c]{@{}l@{}}Interpretation\\ View\end{tabular}}}} & \textbf{Latency}               & 29/25             & 24/37            & 155/140                & -17\%/48\%                                                                       & 546\%/278\%                                                                           & 434\%/460\%                                                                            \\ \cline{2-8} 
\multicolumn{1}{|l|}{}                                                                                        & \textbf{Throughput}            & 33.3/38.3         & 40.9/26.4        & 6.4/7.1                & 23\%/-31\%                                                                       & -84\%/-73\%                                                                           & -81\%/-81\%                                                                            \\ \hline
\multicolumn{1}{|l|}{\multirow{2}{*}{\textbf{\begin{tabular}[c]{@{}l@{}}Assisted\\ Ordering\end{tabular}}}}   & \textbf{Latency}               & 137/3801          & 165/12295        & 886/11444              & 20\%/223\%                                                                       & 437\%/-7\%                                                                            & 547\%/201\%                                                                            \\ \cline{2-8} 
\multicolumn{1}{|l|}{}                                                                                        & \textbf{Throughput}            & 2.3/0.24          & 2.2/0.08         & 0.85/0.085             & -4\%/-67\%                                                                       & -61\%/6\%                                                                             & -63\%/-65\%                                                                            \\ \hline
\end{tabular}
}
\caption{Performance results for sequentially executing 50 times each functionality for 100 and 720 fragments in the database while running inside Docker containers. Results are separated by / in each cell, for instance, by sequentially executing 50 times the \textit{Source Listing} functionality in the monolith we observed an average latency of respectively 17, and 61, milliseconds, where there are respectively 100, and 720, fragments in the database (17/61).}
\label{tab:res10-2-monolith-modular}
\end{table*}

In what concerns the comparison of the performance between the monolith and the modular monolith, it can be observed that the impact on performance of the migration to the modular monolith increases with the number of fragments. In 3 of the 4 functionalities the difference on the performance, between the monolith and the modular monolith, is much higher for 720 fragments than for 100 fragments. This is due to the number of \textit{DTOs} that are generated. Note that for the \textit{Interpretation View} functionality, there is no difference, because it is a functionality that does not depend on the number of fragments in the database. In the \textit{Assisted Ordering} functionality the difference is not so significant as it would be expected, considering the number of operations. This is due to a cache that is being used in both implementations, which reduces the need to transfer \textit{DTOs} between modules, although this is the most computationally demanding functionality.

Additionally, and a little bit surprising, for 100 fragments some functionalities have slightly better performance in the modular monolith. This is due to the fact that one of the modular monolith optimizations is the association of an index to the unique identifiers that are sent to the other modules. The monolith is implemented following an object-oriented approach, where objects are retrieved by searching in an object graph. The optimization in the modular monolith allows a faster retrieval of the domain entities, given their unique identifier, because there are less round trips to the database. This was verified through some test cases, in which this optimization is removed, and the results showed a huge decrease of the performance of the modular monolith. Note that, since the \textit{Interpretation View} functionality does not depend on the number of fragments, it performed better than the monolith even for the tests with 720 fragments.

In what concerns the comparison of the performance between the modular monolith and the microservices architecture, it can be observed that the microservice architecture has a severe negative impact on the performance, both in terms of latency and throughput, independently of the functionality and information in the database. The \textit{Fragment Listing} functionality experienced an extreme negative user experience, reaching a latency average of 5715 ms with a variation of 422\% for 720 fragments in the database, while also experiencing latency variation values of 1089\% for 100 fragments in the database. Similar latency variation values were also observed in the \textit{Source Listing} functionality. These performance values are a consequence of the number of remote invocations necessary to implement the functionality. In each request/response of a remote invocation, the latency values increases due to the network overheads, such as the serialization/deserialization times, that severely affect the performance as the information increases, showing a severe drawback of remote invocations. 

The \textit{Interpretation View} functionality also underwent the effects of the performance drawbacks due to its inter-service communication, presenting a negative impact on the performance. However, the variation was significantly lower than the previously described functionalities and was still capable of providing a good experience to the end-user. This is due to not only a significant difference in the number of remote invocations, but also to the fact that the amount of information sent in the invocation being relatively small, which reduces the network overheads. 

Surprisingly, the microservice architecture did not impact the performance of the \textit{Assisted Ordering} functionality. This is mainly due to two reasons. First, the functionality is computationally demanding, which reduces the impact of distributed communication on the overall performance. Second, the use of caches proved to be effective in optimizing the performance of the functionality. In the modular monolith, a cache that stored the computational results used in each request was already implemented, but, additionally, two more caches were implemented during the development of the \textit{Recommendation} service that stores domain information related to the fragments. Because the functionality repeatedly interacts through the same set of data, these caches reduced the number of remote invocations necessary to perform the functionality.  

During the migration of the modular monolith to the microservices architecture a performance degradation was initially identified due to the high number of fine-grained invocations between the microservices. Therefore, in order to have coarse-grained interactions, it was necessary to increase the amount of domain information preemptively sent in certain \textit{DTOs}, which were frequently utilized together. This replaced remote invocations for local invocations, thus reducing the network overheads and improving the performance. By adding additional information to the \textit{DTOs}, we were able to cache the information and support the composition of several \textit{DTOs} in a single object. With this simple change, the information is now accessible through a single invocation, where before would require several inter-service invocations. However, it is necessary to consider some trade-offs when choosing the information to send, to avoid sending information that is not necessary which will introduce additional overheads. It is necessary to analyse whether the data is frequently used together by the functionalities.

Overall, when comparing the performance of the monolith with the microservices system, and although all the optimizations, the differences are significant for all functionalities, varying the latency between 201\% and 6977\%. This is the consequence of the introduction of additional network overheads that are caused by remote invocations. Note that, a major performance bottleneck of the remote invocations came from the need to serialize and deserialize the \textit{DTOs} on each invocation, due to how it introduces additional latency that becomes more noticeable as the information increases. For instance, we measured and observed that even in a coarse-grained communication, the serialization/deserialization time of the Source Listing functionality corresponded to 82\% of the average latency of the functionality, reaching a serialization time over 3000 ms and a deserialization time of 560 ms. This is a considerable impact on the performance of the functionality, which further increases the latency of the functionality in the modular monolith.

\subsection{Data consistency}

The decentralization of the application data introduces challenges to the architecture concerning its data consistency, due to the transactional behaviour between the different databases. Transactions that span across multiple services cannot implement ACID properties between them.

In the microservices implementation most functionalities are read functionalities, which access one or more microservices to present results to the end user. These functionalities do not create intermediate states, in case of failure, which would generate inconsistent states, due to the lack of isolation. On the other other, functionalities that change data are implemented such that the write occurs in the last local transaction to be executed, and so, they do not introduce inconsistent states. 

However, when the implementation of a functionality implies the asynchronous communication between microservices it introduces eventual consistency between their states. The state of the microservice that receives the event is outdated, compared to the state of the service that publishes it. Their states will be eventually consistent, but, meanwhile, it is possible for a functionality, which accesses both microservices, to observe inconsistent states. This may result, for instance, in the failure of the functionality, if the second microservice gives the functionality an identifier of a domain entity that does not exist in the first microservice.

Most of the events are related with the remove of a domain entity. For instance, upon the removal of a \texttt{Fragment} from the \textit{Text} microservice, the \textit{Virtual} microservice needs to receive the appropriate event to delete any reference to this fragment from the database. This means that it is important to understand the different type of events sent between the microservices, how they affect the data consistency of the different services and how eventual states affect the different functionalities.

\begin{table*}[hbt!]
\resizebox{\textwidth}{!}{%
\begin{tabular}{|c|c|c|c|c|c|c|}
\hline
\textbf{Event Type}                   & \textbf{Publisher}    & \textbf{Subscriber} & \textbf{Impacted Domain Entities}                                               & \textbf{Frequency} & \textbf{Event Trigger}                                              & \textbf{Impact} \\ \hline
\textbf{Fragment-Remove}              & Text                  & Virtual             & VirtualEditionInter                                                             & Low                & \begin{tabular}[c]{@{}c@{}}VEInter-Remove\\ Tag-Remove\end{tabular} & High            \\ \hline
\textbf{ScholarInter-Remove}          & Text                  & Virtual             & VirtualEditionInter                                                             & Low                & \begin{tabular}[c]{@{}c@{}}VEInter-Remove\\ Tag-Remove\end{tabular} & High            \\ \hline
\textbf{SimpleText-Remove}            & Text                  & Virtual             & Annotation                                                                      & Low                & Tag-Remove                                                          & High            \\ \hline
\multirow{3}{*}{\textbf{User-Remove}} & \multirow{3}{*}{User} & Virtual             & \begin{tabular}[c]{@{}c@{}}Member,\\ SelectedBy\\ Tag\\ Annotation\end{tabular} & Low                & Tag-Remove                                                          & High            \\ \cline{3-7} 
                                      &                       & Game                & ClassificationGame                                                              & Low                & -                                                                   & High            \\ \cline{3-7} 
                                      &                       & Recommendation      & RecommendationWeights                                                           & Low                & -                                                                   & High            \\ \hline
\textbf{VirtualEdition-Remove}        & Virtual               & Game                & ClassificationGame                                                              & Medium             & -                                                                   & Low             \\ \hline
\textbf{VirtualEditionInter-Remove}   & Virtual               & Game                & ClassificationGame                                                              & Medium             & -                                                                   & Low             \\ \hline
\textbf{Tag-Remove}                   & Virtual               & Game                & ClassificationGame                                                              & Medium             & -                                                                   & Low             \\ \hline
\textbf{Virtual-Export}               & Virtual               & Game                & -                                                                               & Medium             & -                                                                   & Low             \\ \hline
\end{tabular}
}
\caption{Different event types with the corresponding publisher/subscriber services, impacted information, frequency of occurrence and the impact on the overall information }
\label{tab:event-type}
\end{table*}

Table~\ref{tab:event-type} presents an overview of the different type of events of the LdoD application with the respective publisher/subscriber microservices and impact on data consistency. It can be observed 8 different types of events that affect different microservices and domain entities with different degrees of frequency and impact on the information. Most events are published upon the removal of domain entities that are referenced from microservices that use the modules that contain the domain entities. Additionally, the events can trigger a chain of events which further increases the inconsistency of the application. 

Concerning the impact of the events, the domain entities \texttt{Fragment} and \texttt{ScholarInter} from the \textit{Text} service and \texttt{User} from the \textit{User} service have the highest impact on the consistency, affecting three different databases and multiple domain entities of the \textit{Virtual}, \textit{Game} and \textit{Recommendation} services. In addition, they are also responsible for a chain of events in the \textit{Virtual} service, by triggering three different types of events, \texttt{VirtualEditionInter-Remove}, \texttt{VirtualEdition-Remove}, and \texttt{Tag-Remove}, which further increase the inconsistency in the services. Note that, the impact of multiple databases in this case study is related with the associations removed between the subdomains, which creates indirect relations between the databases that need to be kept consistent through the use of events, because the used services cannot use the services that use them.

The events have a higher impact if they trigger a large number of changes, like the domain entity remove events of the \textit{Text} and \textit{User} services which, fortunately, are the ones that have a lower frequency. On one hand, the archive has a predefined set of fragments that do not change, which means that they are almost like immutable entities, they are loaded once. On the other hand, it is very uncommon to delete the archive users. Therefore, the \textit{Text} and \textit{User} service information remains mostly unchanged throughout the execution of the application under a normal operation, and are only removed under exceptional situations with administrator privileges. This makes the occurrence of these events very rare, which reduces the complexity of managing the consistency. 

Therefore, under normal operation, most of the inconsistencies result from the \textit{Virtual} service subdomain and their associations to the \texttt{ClassficationGame}. The \textit{Virtual} service has a significant number of events that are frequently published because a end-user can interact with its virtual edition, removing some of their entities. On a positive note, the impact of the inconsistent information is mainly focused on the \textit{Game} service and has a low effect on the overall behavior of the application.

In general, the decentralized data approach and eventual consistency achieved through the event-driven asynchronous communication proved to be an efficient alternative to a distributed transactional behaviour between the services in the microservice architecture, where most of the functionalities presented the same behaviour and were not affected by inconsistencies between the databases. This is mainly due to the simple transactional behaviour of the functionalities that do not span across different services, which, in most situations, allows the use of the local transactions and for maintaining the information consistent, thus having an overall low impact.

Finally, to improve the \textit{LdoD Archive} performance, different caches were implemented in the modular and microservice architecture. However, throughout their use in the application, these caches might become inconsistent as the information changes and thus affects the consistency of the application. Therefore, it is also necessary to do a similar evaluation of the caches data consistency and their impact on the application behavior. The analysis resulted in similar conclusions, most of the data in the caches was mostly immutable. Note, however, that the result of this type of analysis depends of the application semantics.

\section{Discussion}
\label{sec:discussion}

The process of modularizing a monolith and migrating it to a microservice architecture requires extensive refactoring of the application and has a significant impact on the performance. The modular monolith and microservice architecture share the modularization requirements to address the decomposition of the domain into the modules/services, while also addressing the granularity of the interactions between the domain entities for performance reasons. Therefore, the modular monolith offers a beneficial groundwork for achieving a microservice architecture. It manages the migration by the stepwise definition of the modularization through well-encapsulated modules, which serve as the foundation of services, and break the migration effort.

Through the evaluation, we addressed some concerns that are often neglected in the literature, which focuses more on technical aspects like communication technology, running environments, and performance benchmarks. In what concerns the migration effort, a significant effort was required for the decomposition of the domain model, but it can be done in the modular monolith migration, and so, free of the complexities of distributed programming. In what concerns the migration from the modular monolith to the microservices architecture, the inter-service communication required changes in the interfaces of each service and the cost is directly related to the size and quality of the REST API. Even though this refactoring is composed of small changes, a poor quality interface can increase the refactoring cost and propagate the changes to the service features. If these constraints are considered when designing the modular interfaces, it will help in the introduction of remote invocations.

Therefore, in the context of a stepwise migration of a monolith into a microservice architecture, the intermediate step of a modular monolith is advantageous, because it fosters separation of concerns, highlighting complexities that might have to be addressed before implementing a microservice architecture, and helps on the decision on how to migrate. Note that, with the modular monolith, the developers can predict the coupling of the services, the expected performance degradation of the communication and decide the type of communication between the services to detect functionalities that can be affected by the lack of ACID transactional behaviour. Additionally, it is possible to foresee the need to redefine a functionality, for instance, due to performance, and proceed to its refactoring in the context of the modular monolith, which is a less complex environment than the microservices system.

On the other hand, we could also observe a consequence of migrating a modular monolith with a significant number of uses relationships like the \textit{LdoD Archive} in terms of coupling between the services. As previously stated, the behaviour of the uses interaction corresponds to a synchronous request/response style of communication between the services in order to maintain the dependencies. However, the synchronous communication results in the coupling of the services being too tight due to the fine-grained behaviour, which becomes problematic, due to the weight of remote invocations, and results in serious performance degradation. This analysis may trigger the need to redefine the functionalities to have an asynchronous behavior.

In what concerns performance, the impact on the performance was thoroughly evaluated. Focusing on the modular monolith, the impact on the performance was related to the amount of information sent between the modules through the \textit{DTOs}, while the number of inter-module invocations did not have a relevant effect. Note that, the modular monolith, when the amount of information transferred between modules is low, can match and even obtain better performance in some cases due to faster accesses to the database through the unique identifiers that were introduced with the modularization. However, contrary to the modular monolith, the performance of the microservice architecture was affected by the number of inter-service invocations. These two factors combined should be avoided when designing the microservice architecture, where there is a performance degradation as the amount of transferred information and the number of invocations increases. 

The impact of performance degradation has to be considered when deciding whether to migrate to a microservices architecture. The latency it adds to functionalities, that are implemented through synchronous invocations of remote transactions, can impair the application usability. Therefore, if the decision is to migrate, it may be necessary to redefine the functionalities that have worse performance, changing the perception the end user has of the application. For instance, the \textit{Fragment Listing} functionality could be redefined to present the results using pagination. This need to change the functionality behavior is often ignored by the literature.

In terms of data consistency, it was necessary to analyse the impact of eventual consistency associated with asynchronous communication. An analysis of the published events, their functionality behavior impact and frequency, is important to identify possible consistency issues. Although it was not the case, these consistency problems may also require a redefinition of the functionalities behavior. Note that, the \textit{LdoD Archive} do not have a significant number of functionalities that do multiple writes, which would require an implementation using sagas~\cite{Garcia-Molina87,Richardson19}. Sagas require some sort of compensation in case of failures, which would significantly increase the migration effort and introduce new challenges. The modular monolith may also help on the analysis of which cases to consider.

The following threats to the validity of this study were identified: (1) it is a single example of a migration; (2) it depends on the technology and programming techniques used in the monolith.

Despite being a single case study, it has some level of complexity, the monolith has 71 domain entities and 81 bidirectional relations, and the literature lacks descriptions of the problems and solutions associated with the migration from monolith to microservices architecture. This study addressed both the migration into a modular monolith and microservice architecture, and provided feedback on challenges faced in the migration that benefits the overall process.

The technology and programming techniques used in the implementation of the monolith follow an object-oriented approach, where the behavior is implemented through fine-grained interactions between objects. A more transaction script~\cite{Fowler02} based implementation of the functionalities may result in different types of problems. The conclusions of this study apply when the monolith is developed using a rich object-oriented domain model. On the other hand, the monolith is implemented using Spring-Boot technology which follows the standards of web application design.

Additionally, the case study presented specific types of functionalities that either requested significant amounts of information or were implemented through fine-grained invocations with a strong synchronous behaviour. Different type of functionalities can present other performance results, but in general most functionalities in the \textit{LdoD Archive} were similar to the selected functionalities. Therefore, a good coverage of the functionalities is provided.

\section{Related Work}
\label{sec:related_work}

There are several challenges when migrating from monolith systems to a microservices architecture~\cite{Francesco18,Kalske18}, such as the development effort to migrate the monolith and the performance impacts, and we can find in the literature the description of the migration of some large monoliths~\cite{Gouigoux17,Bucchiarone18,Barbosa20}.

The migration described by Gouigoux and Tamzalit~\cite{Gouigoux17} discusses the aspects of service migration, deployability and orchestration. It reports an improvement in the performance, but does not describe details of the refactorings and optimizations done. Bucchiarone et al~\cite{Bucchiarone18} describe the lessons learned from their migration of a monolith in the banking domain, focusing on the benefits and challenges of the new system, but do not discuss the migration effort and the impact on performance. Barbosa and Maia~\cite{Barbosa20} discuss the migration of a large monolith to a microservices architecture, where the monolith business logic is implemented using stored procedures, and they focus on the process of identifying microservices. Megargel et al~\cite{megargel2020migrating} provide a practice-based view and a methodology to transition from a monolith application into a cloud-based microservice architecture. Despite addressing the different technology aspects and describing the migration process, the migration effort was not addressed, although they did observe benefits from the migration, including on the performance. On the other hand, an opposite point of view was presented by Mendonça~\cite{mendonccamonolith} who discussed the decision to revert the microservice architecture back into a monolith, focusing on the burdens from the microservice architecture and the lessons learned. The researchers also addressed the cost of different aspects of the microservice architecture such as the scalability, security isolation and deployment, but, even in this case, migration effort and performance are not addressed. 

Therefore, it is necessary to have more case studies that describe the migration efforts and the architectural trade-offs, besides the advantages, and drawbacks, of the final product. Due to the technological complexity, it is beneficial that these studies follow a separation of concerns approach, such that the analysis can be done according to levels of complexity. In this paper we contribute by describing the refactoring, and related effort, associated with the migration of a large monolith to a microservices architecture, using a modular monolith as an intermediate step.

On the other hand, there is work on impact that the migration of a monolith to a microservices architecture has on performance, although it follows different perspectives.

Ueda et al.\cite{Ueda16} compare the performance of microservices with monolith architecture to conclude that the performance gap increases with the granularity of the microservices, where the monolith performs better. Villamizar et al.~\cite{Villamizar15} show different results, concluding that in some situations the performance is better in the microservices context and that it reduces the infrastructure costs, but request time increases in microservices due to the gateway overhead. Al-Debagy and Martinek~\cite{Al-Debagy18} conclude that they have similar performance values for average load, and the monolith performs better for a small load. In a second scenario the monolith has better throughput, but similar latency, when the system was stressed in terms of simultaneous requests. Bjørndal et al.~\cite{Bjorndal20} benchmark a library system, that has four use cases and considers synchronous and asynchronous relations between microservices. They observe that the monolith performs better except for scalability. Therefore, they identify the need to carefully design the microservices, in order to reduce the communication between them to a minimum, and conclude that it would be interesting to apply these measures in systems that are closer to the kind of systems used by companies. Guamán et al.~\cite{guaman2018performance} designed and implemented a multi-stage architectural migration into a microservice architecture and compared the performance of the monolith and the intermediary stages to the microservice architecture. They conclude that the microservice architecture has worse latency. Flygare et al.~\cite{flygare2017performance} found that in their case study the monolith performed better in terms of latency and throughput while consuming less resources than the microservice architecture. In addition, they also observed some throughput benefits in running the microservice in a cluster instead of running in a single computer.

Some other perspectives compare the performance of monolith and microservices systems in terms of the distributed architecture of the solution, such as master-slave~\cite{Amaral15}, the characteristics of the running environment, whether it uses containers or virtual machines~\cite{Joy15}, the particular technology used, such as different service discovery technologies~\cite{Al-Debagy18}, or other microservices deployment aspects~\cite{Tapia20}. A major aspect of the performance is the type of inter-service communication and the technology used. Hong et al.~\cite{hong2018performance} compare the performance of synchronous and asynchronous communication and concluded that the asynchronous approach offered a more stable performance overall, but with a lower response request performance. Fernandes et al.~\cite{fernandes} presented similar results with the asynchronous communication outperforming the REST communication in performance and data loss prevention on a large data context. Shafabakhsh et al.~\cite{shafabakhsh2020evaluating} leverage on Fernandes et al.~\cite{fernandes} research and conclude that there is a benefit of synchronous communication under small loads.

These results show that an asynchronous communication tends to be more suitable to a microservice architecture. However, a behaviour suited synchronous communication like the REST API is more accessible to the migration from a modular monolith and reduces the complexity of redesigning the functionalities into an asynchronous behaviour. Our approach allows to separate concerns when measuring the performance impact of the migration, and led us to conclude that it is already significant when migrating to a modular monolith, in the absence of distributed communication and microservices implementation technologies. And, contrarily to some of the related work, in which there is no redefinition of the functionalities of the monolith, we highlight that such redefinition may be required due, for instance, to performance degradation.

\section{Conclusion}
\label{sec:conclusion}

With this work, we described the migration from a large object-oriented monolith into a microservice architecture using a modular monolith as a middle stage, while analysing the migration effort to achieve both, the modular and microservice architectures, as well as the overall impact on the performance.

The results show that the modular monolith can be used as an intermediate artifact that facilitates the migration into the microservice architecture, while providing a more agile software development environment before tackling the challenges of a microservice architecture. In addition, the modular monolith also helped in addressing concerns regarding the inter-service communication, eventual consistency and performance optimizations before the second migration step. 

The migration into a microservice architecture revealed a significant impact on the application for both refactoring and performance. Most of the migration cost was connected to the modules and the interfaces to implement services with the desired inter-service communication, in which the quality of the modules and interfaces has a significant impact on the cost. On the other hand, a serious consequence of the migration was the large impact on performance associated with the inter-service communication, which required functionalities to be reimplemented to have coarse-grained interactions. 

Overall, the migration to a microservices architecture presented several challenges with different levels of impact on the migration effort, performance, and data consistency, which highly depended on the application structure and semantics. In the \textit{LdoD Archive}, the migration presented a serious impact on the performance due to the network overheads that proved to be far too high compared to the previous architectures, but on the other hand, it offered a more scalable and manageable architecture.

Although we describe a single case study of migration, it is of a system that is in production, and implemented using the best practices for the development of web applications. Additionally, our stepwise approach seems suitable for further studies, because it fosters a separation of concerns that helps on the controlled introduction of complexity, and on the analysis of the results.

The code repository is publicly available in a GitHub repository\footnote{\url{https://github.com/socialsoftware/edition}}, in which the branch \textit{master} contains the monolith and branches \textit{modular-monolith} and \textit{microservices} contain the result of the respective migrations. 

\section*{Acknowledgement}
This work was supported by national funds through FCT, Fundação para a Ciência e a Tecnologia, under project UIDB/50021/2020.

\bibliographystyle{./bibliography/IEEEtran}

\end{document}